\documentclass[9pt,twocolumn,twoside]{osajnl}

\journal{josaa}
\setboolean{shortarticle}{true}

\title{Modeling circulating cavity fields using the discrete linear canonical transform}

\author[1,2,*]{A. A. Ciobanu}
\author[1,2]{D. D. Brown}
\author[1,2]{P. J. Veitch}
\author[1,2]{D. J. Ottaway}

\affil[1]{Department of Physics, School of Physical Sciences and The Institute of Photonics and Advanced Sensing (IPAS), The University of Adelaide, SA, 5005, Australia}
\affil[2]{Australian Research Council Centre of Excellence for Gravitational Wave Discovery (OzGrav)}
\affil[*]{Corresponding author: alexei.ciobanu@adelaide.edu.au}

\setboolean{displaycopyright}{true}

\usepackage{ parskip }
\usepackage{ amssymb }
\usepackage{ mathtools } 
\usepackage{ graphicx }
\usepackage{ xcolor }
\usepackage{ hyperref }
\usepackage{ tabularx }
\usepackage{ gensymb }
\usepackage{ lipsum }
\usepackage{ cleveref }

\definecolor{red}{RGB}{255,0,0}
\definecolor{green}{RGB}{0,128,0}
\definecolor{orange}{RGB}{255,128,0}
\definecolor{blue}{RGB}{0,128,255}

\def\comments_on{1}
\def\paragraphs_on{1}

\newcommand{\comment}[1]{\if\comments_on1
#1
\fi}

\newcommand{\tem}[1]{HG$_{#1}$}

\newcommand{\iu}{\mathrm{i}\mkern1mu}

\newcommand{\lct}[1]{\mathcal{L}_{#1}}
\newcommand{\dlct}[1]{\mathbf{L}_{#1}}

\makeatletter
\newcommand{\pushright}[1]{\ifmeasuring@#1\else\omit\hfill$\displaystyle#1$\fi\ignorespaces}
\newcommand{\pushleft}[1]{\ifmeasuring@#1\else\omit$\displaystyle#1$\hfill\fi\ignorespaces}
\makeatother

\LetLtxMacro{\oldparagraph}{\paragraph}
\renewcommand{\paragraph}[1]{\if\paragraphs_on1
\oldparagraph{#1}\mbox{}\newline
\fi}

\begin{abstract} 

Fabry-Perot cavities are central to many optical measurement systems. In high precision experiments, such as aLIGO and AdV, coupled cavities are often required leading to complex optical dynamics, particularly when optical imperfections are considered. We show, for the first time, that discrete LCTs can be used to compute circulating optical fields for cavities in which the optics have arbitrary apertures, reflectance and transmittance profiles, and shape. We compare the predictions of LCT models with those of alternative methods. To further highlight the utility of the LCT, we present a case study of point absorbers on the aLIGO mirrors and compare with recently published results.

\end{abstract}

\begin{document}

\maketitle


\section{Introduction}

Optical cavities, such as Fabry-Perot interferometers are commonly used in precision optical experiments. The circulating field of any single geometrically stable cavity is a sum of the Hermite-Gauss (HG) modes of the cavity, which are given by analytical expressions~\cite{kogelnik_laser_1966,siegman_lasers_1986}. 

HG decomposition can accurately model the steady state field that exists in multiple coupled cavities and is extensively used for optical modelling for the advanced gravitational wave detectors ~\cite{brown_pykat_2020}.
This method however, is inefficient for modelling high spatial-frequency features, such as finite mirror apertures and small scale aberrations such as point absorbers~\cite{brooks_point_2021}. This is because of the prohibitively large number of HG modes needed to accurately model them~\cite{bond_interferometer_2016}.


The alternative to HG decomposition is to consider the complex beam amplitude sampled on a uniform cartesian grid.
In such a model the spaces in an optical system are represented by diffraction integrals and the optical elements (e.g. mirrors, and lenses) are approximated with thin phase plates. 
The accuracy of the model is determined by the resolution of the sampled grid.
This allows non gaussian features to be accurately modelled to the resolution limit of the sampled grid.

In the cartesian basis the diffraction integral becomes a linear operator that couples all points in the input plane to all points in the output plane.
This causes the issue that in two dimensions the size of the linear operator grows as the square of the number of grid points, quickly becoming too large for practical simulations due to memory limitations.
A common optimization is to perform the diffraction integrals in the Fourier domain where the convolution in the diffraction integral becomes multiplication in the Fourier domain, and hence cheaper to compute.
This optimization has led to the use of iterative algorithms for solving for circulating fields in the cartesian basis.


The first known usage of an iterative algorithm for solving steady state circulating fields is the \textit{Fox-Li method} demonstrated in 1961~\cite{fox_resonant_1961}.
The method involves taking a guess at a circulating field and repeatedly computing round trip propagations inside the cavity until the circulating field converges to a steady state,  with the number of iterations typically scaling with the finesse of the cavity~\cite{vinet_high_1992}.
The speed of each iteration is typically limited by the diffraction integral.
Fast Fourier transforms (FFTs) can be used to efficiently compute the diffraction integrals, which has in turn led to this class of models being colloquially called \textit{FFT~models}.
The number of iterations is reduced with a better initial guess of the circulating field or by accelerating the rate of convergence by using a modified iteration scheme~\cite{saha_fast_1997,bochner_grid-based_2003, day_accelerated_2014}.
The Fox-Li method has also been used in modelling eigenmodes of unstable resonators~\cite{siegman_laser_2000-1,siegman_laser_2000}.

FFT models can be used to model more complicated optical systems involving multiple coupled cavities, such as the Fabry-Perot power recycled Michelson interferometers (FPPRMI) used in Gravitational Wave (GW) detectors as modeled by Bochner et al. in 2003~\cite{bochner_grid-based_2003}.
Bochner et al. additionally state that FFT models are difficult to implement in code and that accelerating the rate of convergence of their FFT models often leads to instability---which was also found to be an issue by Day et al.~\cite{day_accelerated_2014}.
A 2017 review of unstable resonator eigenmode modelling by New~\cite{new_computing_2017} showed that the Fox-Li method can be outperformed by general purpose linear algebra algorithms available in MATLAB in cases where the sampling grid resolution is not larger than $512\times512$.



Linear Canonical Transforms (LCT) provides an alternative modeling formalism to modal and FFT models.
The LCT was first introduced to the optical modelling community in 1970 by Collins~\cite{collins_lens-system_1970} and has been widely used to derive beam propagation formulas and model single pass optical systems.
Recently the LCT has been applied in modelling 1D unstable cavity eigenmodes~\cite{wang_eigenvector_2012}, and propagation in Herriott cells~\cite{dahlen_modeling_2017}.
Over the past 20 years there have been many efforts to refine discrete approximations of the LCT across multiple disciplines~\cite{soo-chang_pei_closed-form_2000,pei_fast_2016,healy_simulating_2018,koc_discrete_2019}.


In this paper we apply the discrete LCT (DLCT) for modeling circulating fields in resonant cavities, for the first time to the best of our knowledge.
A particular benefit of the LCTs is that it enables us to use highly optimized general purpose linear algebra algorithms instead of the Fox-Li method.
The paper is laid out as follows: in section~\ref{sec:continuous_LCT} we define the continuous LCT and some of its properties. 
Section~\ref{sec:LCT_wave_propagation} covers a brief review of the application of the LCT in optical propagation in ABCD optical systems.
Section~\ref{sec:cavity_LCT} demonstrates a new application of the LCT to calculate the circulating field of single linear and ring cavities, as well as linear cavities with finite apertures and mirror deformations including point absorbers.
Appendix~\ref{sec:metaplectic_sign} also contains a discussion on the metaplectic nature of the LCT, which is used to derive new expressions for accumulated Gouy phase of HG modes in ABCD optical systems.

\section{Linear Canonical Transform}\label{sec:continuous_LCT}

The history of the linear canonical transform (LCT) can be traced back to two independent origins ~\cite{liberman_structural_2017,healy_development_2016}; one in optics~\cite{collins_lens-system_1970}, and one in quantum mechanics~\cite{moshinsky_linear_1971}. 
In optics alone the LCT has appeared under a number of other names such as: the generalized Huygens' integral~\cite{siegman_20_1986}, and the affine Fourier transform~\cite{abe_optical_1994}.
More abstractly the LCT is a faithful representation of the metaplectic group acting on phase space~\cite{bacry_metaplectic_1981, wolf_geometric_2004}.

The LCT is a paraxial diffraction integral that models the propagation of an arbitrary electric field through an optical system that is represented by an ABCD matrix~\cite{siegman_20_1986}. 
Formally, the LCT is a family of integral transforms, which include the Fourier and fractional Fourier transforms, as well as Laplace transforms, and the Fresnel integral.
 Any particular LCT can be parameterized up to an overall sign by four complex numbers denoted $A,B,C,D$, and one constraint $AD - BC = 1$~\cite{ozaktas_fractional_2001,healy_linear_2016}.

\subsection{Continuous LCT}

The continuous LCT, $g(x)$, of a function $f(x)$ is defined as
\begin{align}
	g(x_2) &= \int_{-\infty}^\infty \lct{}(x_1, x_2) f(x_1) dx_1 \label{eq:active_LCT_def}
\end{align}
where $\lct{}$ is a linear operator and the kernel of the LCT integral transform
\begin{align}
    \lct{}(x_1, x_2) &= \begin{dcases}
        \sqrt{\frac{\iu}{ B \lambda}} \times \exp\! \left[ \frac{-\iu \pi}{B \lambda} \left( A x_1^2 - 2 x_1 x_2 + D x_2^2 \right) \right] & \\
	 {\sqrt{D}} \times \exp\! \left[\frac{-\iu \pi C}{\lambda D} x_2^2 \right] \delta(x_1 - D x_2) \qquad \quad \text{if } B=0 
	\end{dcases}
	\label{eq:continuous_LCT_def}
\end{align}
where $\lambda$ is the optical wavelength, $\{A,B,C,D\}$ are the LCT parameters, and $\delta(x)$ is the Dirac delta function. 

\subsection{Discrete LCT}
A discrete LCT can be obtained by taking $N$ samples of \eqref{eq:active_LCT_def} at regular intervals $\Delta x$
\begin{align}
    g(x_k) =  \Delta x \times \sum_{j=1}^N \lct{}(x_j, x_k) f(x_j)
    \label{eq:active_DLCT_def}
\end{align}
where $\Delta x = \mathrm{x}_{i+1} - \mathrm{x_i}$, and $f(x) \to 0$ for $x<x_1$ and $x>x_N$.

We can simplify the notation by introducing the following N-vector $\mathbf{x}$ 
\begin{align}
    \mathbf{x} = \left[\mathrm{x}_1, \mathrm{x}_2, \ldots ,\mathrm{x}_{N-1}, \mathrm{x}_{N}\right]^T.
\end{align}
Then \eqref{eq:active_DLCT_def} can be rewritten as a matrix-vector product
\begin{align}
    \mathbf{g} = \Delta x \times \dlct{} \mathbf{f}
\end{align}
where $\mathbf{f} = f(\mathbf{x})$, $\mathbf{g} = g(\mathbf{x})$ are N-vectors, and $\dlct{}$ is an $N \times N$ matrix. The matrix $\dlct{}$ is given by
\begin{align}
    \dlct{} = \sqrt{\frac{\iu}{ B \lambda}} \times \exp_\circ\left[ \frac{-\iu \pi}{B \lambda} \left( A \mathbf{X_1}^2 - 2 \mathbf{x} \mathbf{x}^T + D \mathbf{X_2}^2 \right) \right]
\end{align}
where $B \neq 0$, $\exp_\circ[\mathbf{X}]$ is an elment-by-element exponentiation,
\begin{align}
    \mathbf{X_1}^2 = \begin{bmatrix}
        x_1^2 & \ldots & x_N^2\\
        \vdots & \ddots & \vdots\\
        x_1^2 & \ldots & x_N^2
    \end{bmatrix} &, & \mathbf{X_2}^2 = \begin{bmatrix}
        x_1^2 & \ldots & x_1^2\\
        \vdots & \ddots & \vdots\\
        x_N^2 & \ldots & x_N^2
    \end{bmatrix},
\end{align}
and $\mathbf{x} \mathbf{x}^T = \mathbf{x} \otimes \mathbf{x}$ is an outer product.

This DLCT implementation is sometimes called the ``direct'' implementation~\cite{healy_simulating_2018}.
It accurately approximates the continuous LCT except when $|B| \ll 1$, for which the magnitude of the exponent in \eqref{eq:continuous_LCT_def} become large and thus a higher resolution array of samples is required to avoid aliasing in the kernel $\dlct{}$. 

A number of alternative DLCT implementations have been constructed over the years, most of which are well-behaved for $|B| \ll 1$~\cite{pei_fast_2016, koc_discrete_2019}.
However, it appears there is no single DLCT implementation that has a consistently lower approximation error for all possible LCT parameters and input functions.

For modelling geometrically stable resonant optical cavities we only consider LCTs for geometrically stable cavities, where $|A + D| < 2$, and $|B|$ is not close to zero, for which the direct implementation in \eqref{eq:active_DLCT_def} performs adequately.

\subsection{LCT Composition}

Composition is perhaps the most powerful property when working with LCTs.
For optical models the composition property allows for any arbitrary ABCD optical system to be accurately modelled with a single LCT. 
This drastically reduces the amount of computation compared to equivalent FFT models, which require each optical component in a system to be modelled individually.

We adopt the standard convention of packaging the $\{A,B,C,D\}$ parameters into a matrix \hbox{$\mathbf{M} = \Big[\begin{smallmatrix} A & B\\C & D\end{smallmatrix}\Big]$}.
These matrices use the `reduced'' ray slope method defined in ~\cite{siegman_15_1986}. 
They are identical to the standard ABCD matrices~\cite{kogelnik_laser_1966}, except for the case where a refractive index change occurs.

The composition property states that composing two LCTs is the same as single LCT (up to a sign difference) whose ABCD matrix is given by the matrix product of the ABCD matrices of the composed LCTs~\cite{littlejohn_semiclassical_1986}
\begin{align}
	\lct{\mathbf{M_2}} \lct{\mathbf{M_1}} &= \sigma \lct{\mathbf{M_2 M_1}}, \label{eq:lct_composition}
\end{align}
where $\sigma \in \{-1,1\}$ is the metaplectic sign.
In general, when composing ABCD matrices it is necessary to keep track of the metaplectic sign $\sigma$.
Failing to do so may introduce erroneous minus signs in the calculation of the accumulated Gouy phase.
Further discussions on the metaplectic sign, as well as instructions on how to compute it are presented in appendix~\ref{sec:metaplectic_sign}.

Unfortunately, all known DLCT implementations only satisfy composition approximately~\cite{pei_two-dimensional_2016, koc_discrete_2019}.
\begin{align}
	\dlct{\mathbf{M_2}} \dlct{\mathbf{M_1}} &\approx \sigma \, \dlct{\mathbf{M_2 M_1}} \label{eq:dlct_composition}
\end{align}
Exact composition for specific subsets of LCT parameters can be recovered in some DLCT implementations~\cite{zhao_constraints_2015}.
This approximation error can be made arbitrarily small by using higher resolution DLCT matrices $\dlct{\mathbf{M}}$.
This is similar to how the accuracy of an FFT based model scales with grid resolution.



\section{Optical propagation using the LCT} \label{sec:LCT_wave_propagation}
Here we review a well known application of the LCT: the propagation of paraxial electric fields through an ABCD optical system.
We discuss both analytical solutions of the LCT and its discrete approximations to 1D and separable 2D systems.

\subsection{1D propagation}

An 1D paraxial electric field $E_\text{in}(x)$ at an input plane can be propagated through a paraxial optical system represented by an ABCD matrix $\mathbf{M}$ using \eqref{eq:active_LCT_def}
\begin{align}
	E_\text{out}(x') = \int_{-\infty}^{\infty} \lct{\mathbf{M}}(x,x') E_\text{in}(x) dx
	\label{eq:1D_LCT}
\end{align}
where $E_\text{out}(x)$ is the electric field at the output plane.

In the discrete case, the $N$ samples of the output electric field $\mathbf{u}_{\text{out}}$ can be calculated by matrix multiplying an N-vector representation of the input electric field $\mathbf{u}_{\text{out}}$ by $N \times N$ DLCT matrix $\dlct{\mathbf{M}}$
\begin{align}
	\mathbf{u}_{\text{out}} = \dlct{\mathbf{M}} \mathbf{u}_{\text{in}}.
	\label{eq:1D_DLCT}
\end{align}

\begin{figure}[ht]
	\centering
    \includegraphics[width=\linewidth]{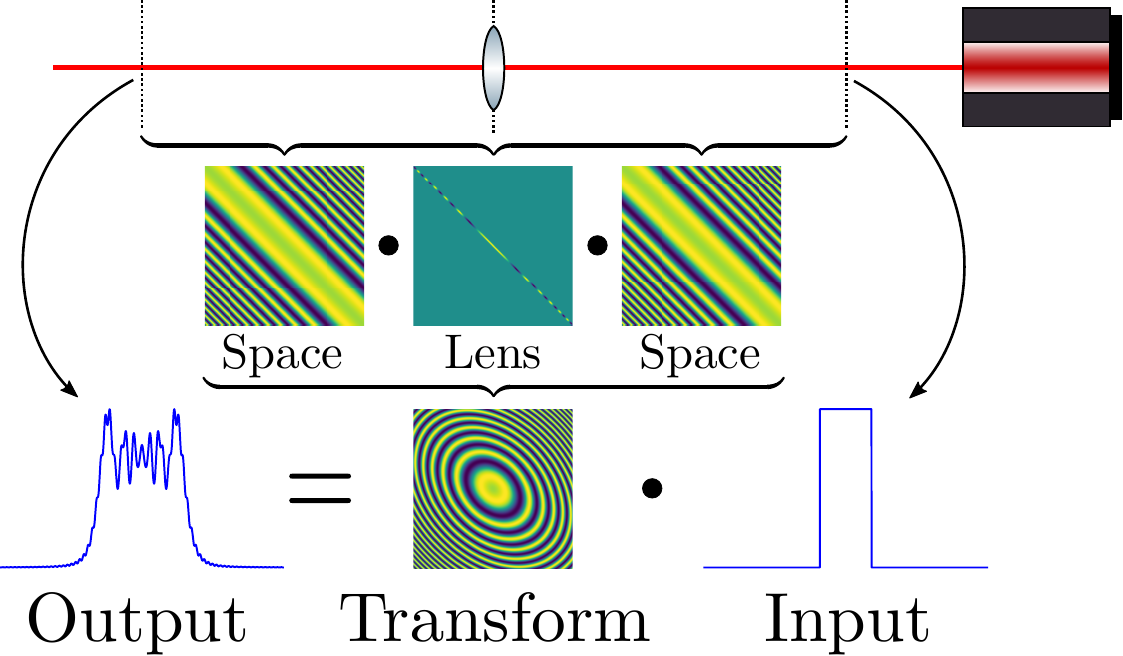}
    \caption{A visualization of a 1D DLCT where the electric field propagates from right-to-left to maintain consistency with the matrix-vector multiplication in the bottom row.
    The optical system consists of 3 ABCD matrices, each of which yields a DLCT kernel, the real part of which is shown.
    The overall kernel is determined using \eqref{eq:dlct_composition}, and used to propagate a tophat beam from the input plane to the output plane using \eqref{eq:1D_DLCT}.
    }
    \label{fig:DLCT_workflow}
\end{figure}

The DLCT provides a simple framework for viewing optical propagation as a product of matrices, each corresponding to a basic optical component, to yield a single DLCT matrix for the overall optical system as illustrated in figure~\ref{fig:DLCT_workflow}.

\subsection{Separable 2D propagation}

For electric fields propagating through a separable 2D optical system 
 \begin{align}
     E_\text{out}(x', y') = \int_{-\infty}^\infty \lct{\mathbf{M}_y} (y,y') \left(\int_{-\infty}^\infty \lct{\mathbf{M}_x} (x, x') E_\text{in}(x,y) dx \right) dy \label{eq:2d_separable_lct}
 \end{align}


In the discrete case, we can approximate the $N \times M$ samples of the output electric field $\mathbf{U}_\text{out}$ using
\begin{align}
    \mathbf{U}_\text{out} = \Delta x \Delta y \times \dlct{\mathbf{M}_y} \mathbf{U}_\text{in} \big(\dlct{\mathbf{M}_x}\big)^T \label{eq:2d_separable_dlct}
\end{align}
where $\dlct{\mathbf{M}_y}$ is an $N \times N$ 1D DLCT matrix for the y-axis and $\big(\dlct{\mathbf{M}_x}\big)^T$ is the transpose of an $M \times M$ 1D DLCT matrix for the x-axis.
This corresponds to an optical system that can be described by two $2 \times 2$ ABCD matrices $\mathbf{M}_x$ and ${\mathbf{M}_y}$, one for each axis.

In the case where the kernel is not separable \eqref{eq:continuous_LCT_def} needs to be modified to support $4 \times 4$ ABCD matrices to form the 2D nonseparable LCT (NsLCT) \cite{pei_two-dimensional_2016, zhao_fast_2019}.
Throughout the rest of this paper we will only consider separable 2D LCTs.

\section{Circulating field in ideal resonant cavities}\label{sec:cavity_LCT}

\subsection{Linear cavities}\label{sec:linear_cavity}

The circulating field in a stable optical cavity is determined by generalizing the strategy used by Siegman~\cite{siegman_11_1986} for plane waves.
\begin{figure}[!ht]
	\centering
    \includegraphics[width=\linewidth]{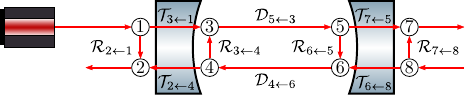}
    \caption{A directed network graph of a Fabry-Perot cavity. 
    Each node corresponds to a distinct electric field, and each edge/red arrow corresponds to a linear operator that transforms a field from the input to the output node.
    Multiple edges leading into a single node are summed together to obtain the field at that node.}
    \label{fig:cavity_graph}
\end{figure}

For the cavity shown in figure~\ref{fig:cavity_graph} the circulating field at node~3 is given by
\begin{equation}
    E_3 = \mathcal{R}_{3\gets4} \mathcal{D}_{4\gets6} \mathcal{R}_{6\gets5} \mathcal{D}_{5\gets3} E_3 + \mathcal{T}_{3 \gets 1} E_1,
    \label{eq:cavity_graph_circulating_field}
\end{equation}
where we assume $E_8 = 0$. The definition of symbols that convert \eqref{eq:cavity_graph_circulating_field} into an LCT model for an ideal cavity is given in table~\ref{tab:linear_cavity_operators}, where $r_1, r_2, \mathbf{M}_{R_1}, \mathbf{M}_{R_2}$ are amplitude reflection coefficients and ABCD matrices of the input and end mirror respectively, $t_1$ and $\mathbf{M}_{T_1}$ is the amplitude transmission coefficient, and ABCD matrix for the input mirror, $\mathbf{M}_{d}$ is the free-space propagation ABCD matrix for the distance between the mirrors, and $\phi = 2 k d$ is the accumulated round-trip plane-wave phase. 

\begin{table}
    \centering
    \begin{tabular}{|r|r|r|r|}
        \hline
        Edge & Analytical & Discrete 1D & Discrete 2D \\
        \hline
        \rule{0pt}{4ex} $\mathcal{T}_{3\gets1} [\mathbf{X}]$ & 
        $\iu t_1 \lct{\mathbf{M}_{T_1}}[\mathbf{X}]$ & 
        $\iu t_1 \dlct{\mathbf{M}_{T_1}}\mathbf{X}$ &
        $\iu t_1 \dlct{\mathbf{M}_{T_{1y}}}\mathbf{X} \left(\dlct{\mathbf{M}_{T_{1x}}}\right)^T$ \\
        $\mathcal{R}_{3\gets4} [\mathbf{X}]$ & 
        $r_1 \lct{\mathbf{M}_{R_1}}[\mathbf{X}]$ & 
        $r_1 \dlct{\mathbf{M}_{R_1}}\mathbf{X}$ & 
        $r_1 \dlct{\mathbf{M}_{R_{1y}}}\mathbf{X} \left(\dlct{\mathbf{M}_{R_{1x}}}\right)^T$ \\
        $\mathcal{R}_{6\gets5} [\mathbf{X}]$ & 
        $r_2 \lct{\mathbf{M}_{R_2}}[\mathbf{X}]$ & 
        $r_2 \dlct{\mathbf{M}_{R_2}}\mathbf{X}$ &
        $r_2 \dlct{\mathbf{M}_{R_{2y}}}\mathbf{X} \left(\dlct{\mathbf{M}_{R_{2x}}}\right)^T$ \\
        $\mathcal{D}_{4\gets6} [\mathbf{X}]$ & 
        $e^{\iu \phi/2} \lct{\mathbf{M}_{d}}[\mathbf{X}]$ & 
        $e^{\iu \phi/2} \dlct{\mathbf{M}_{d}}\mathbf{X}$ &
        $e^{\iu \phi/2}  \dlct{\mathbf{M}_{d}}\mathbf{X} \left(\dlct{\mathbf{M}_{d}}\right)^T$ \\
        $\mathcal{D}_{6\gets5} [\mathbf{X}]$ & 
        $e^{\iu \phi/2} \lct{\mathbf{M}_{d}}[\mathbf{X}]$ &
        $e^{\iu \phi/2} \dlct{\mathbf{M}_{d}}\mathbf{X}$ & 
        $e^{\iu \phi/2}  \dlct{\mathbf{M}_{d}}\mathbf{X} \left(\dlct{\mathbf{M}_{d}}\right)^T$\\[2pt]
        \hline
    \end{tabular}
    \caption{Definition of symbols in \eqref{eq:cavity_graph_circulating_field} for modelling an ideal linear cavity in an analytical, 1D and 2D discrete LCT models.
    The symbol $\mathbf{X}$ is either an analytical expression, a vector, or a matrix respectively.}
    \label{tab:linear_cavity_operators}
\end{table}

Substituting in the LCT operators gives
\begin{align}
	E_\text{circ} = e^{\iu \phi} r_1 r_2 \lct{\mathbf{M}_{R_1}} \lct{\mathbf{M}_{d}} \lct{\mathbf{M}_{R_2}} \lct{\mathbf{M}_{d}} E_\text{circ} + \iu t_1 \lct{\mathbf{M}_{T_1}} E_\text{inc}.
\end{align}
We can use the LCT composition property to simplify the sequence of LCTs $\lct{\mathbf{M}_{R_1}} \lct{\mathbf{M}_{d}} \lct{\mathbf{M}_{R_2}} \lct{\mathbf{M}_{d}}$ into a single round-trip LCT $\lct{\mathbf{M}_\text{RT}}$, and thus
\begin{align}
	E_\text{circ} = e^{\iu \phi} r_1 r_2 \lct{\mathbf{M}_\text{RT}} E_\text{circ} + \iu t_1 \lct{\mathbf{M}_{T_1} } E_\text{inc}
    \label{eq:circ_composed_lct}
\end{align}
Rearranging for $E_\text{circ}$ gives us
\begin{align}
	E_\text{circ} = \iu t_1 \Big(\mathcal{I} - e^{\iu \phi} r_1 r_2 \lct{\mathbf{M}_\text{RT}} \Big)^{-1} \lct{\mathbf{M}_{T_1} } E_\text{inc} 
	\label{eq:lct_circulating_field}
\end{align}
where $\mathcal{I}$ is the identity operator ($\mathcal{I} x = x$). 

Unfortunately, the \hbox{$\big( \mathcal{I} - e^{\iu \phi} r_1 r_2 \lct{\mathbf{M}_\text{RT}} \big)$} operator has no known closed form inverse, but it can be approximated with discrete numerical methods.
Rewriting \eqref{eq:lct_circulating_field} in the 1D discrete case yields the following matrix-vector equation
\begin{align}
    \mathbf{u}_\text{circ} = \iu t_1 \Big(\mathbf{I} - e^{\iu \phi} r_1 r_2 \dlct{\mathbf{M}_\text{RT}} \Big)^{-1} \dlct{\mathbf{M}_{T_1} } \mathbf{u}_{\text{inc}} \label{eq:dlct_circulating_field}
\end{align}
where $\mathbf{I}$ is an $N \times N$ identity matrix. 
\eqref{eq:dlct_circulating_field} can be trivially computed with a matrix inverse algorithm.

For 2D the solution is more complicated. Starting from \eqref{eq:circ_composed_lct} the $N \times N$ 2D discrete circulating field $\mathbf{U}_\text{circ}$ is given by
\begin{align}
    \mathbf{U}_\text{circ} - e^{\iu \phi} r_1 r_2 \dlct{\mathbf{M}_{\text{RT}_y}} \mathbf{U}_{\text{circ}} (\dlct{\mathbf{M}_{\text{RT}_x}})^T = \iu t_1  \dlct{\mathbf{M}_{T_{1y}}} \mathbf{U}_{\text{inc}} (\dlct{\mathbf{M}_{T_{1x}}})^T. \label{eq:as2}
\end{align}
\eqref{eq:as2} is of the form where the left-hand-side (LHS) consists of purely linear operations on the unknown $\mathbf{U}_\text{circ}$ and the right-hand-side (RHS) contains only known quantities.
Therefore, \eqref{eq:as2} is equivalent to an $N^2 \times N^2$ system of linear equations given by $\mathbf{Ax} = \mathbf{b}$.
However, the $\mathbf{Ax} = \mathbf{b}$ form is not practical due to memory limitations arising from the $N^2 \times N^2$ dimensionality.

\eqref{eq:as2} can be rewritten as a Sylvester equation $\mathbf{AX} + \mathbf{XB} = \mathbf{Q}$, where $\mathbf{X}$ is unknown, and $\mathbf{A}, \mathbf{B}, \mathbf{Q}$ are known $N \times N$ matrices, which can be solved efficiently by the Bartels-Stewart algorithm~\cite{bartels_solution_1972} in LAPACK~\cite{demmel_lapack_1989}.
Alternatively \eqref{eq:as2} can be solved using iterative sparse linear solvers such as GMRES~\cite{saad_gmres_1986}, which can also be used to model cavities with mirror surface imperfections and is discussed in more detail in section~\ref{sec:apertured_cavity}.

\subsection{Ring cavities}\label{sec:ring_cavity}

The approach to modelling ring cavities is similar to the linear cavity except that the reflection operators have to be modified to include a parity transformation for the reflected electric field.
This parity transformation is also present in linear cavities, however is often neglected as the parity from each mirror cancels in the round trip.

The triangular cavity is an interesting example as it is the simplest cavity with an odd number of mirrors, which has an overall parity in the round trip and hence splits the resonances of its horizontal and vertical eigenmodes~\cite{collins_modes_1964,collins_analysis_1964,moller_fabry-perot_1998}.
Typically, this phenomenon is modelled by explicitly counting the number of mirrors and manually applying a minus sign in specific equations to produce the desired transverse mode splitting.

For both LCT and FFT models the transverse mode splitting in triangular cavities is generated by parity operators $\mathcal{P}$ that swap the left and right side of the electric field on reflection $\mathcal{P}[E(x, y)] = E(-x, y)$ to maintain a consistent coordinate system~\cite{siegman_15_1986}.
We can identify the mirror counting behavior by noting that the parity operator commutes with the LCT $\mathcal{P} \lct{\mathbf{M}} = \lct{\mathbf{M}} \mathcal{P}$.
The parity operators can thus all be commuted to one side where we only need to consider whether there are an odd or even number of parity operators (i.e. odd or even number of mirrors) because the parity operator is its own inverse $\mathcal{P}\mathcal{P} = \mathcal{I}$.
In the case of a round trip in a triangular cavity we are then left with a single overall parity operator $\mathcal{P}$, which flips the sign of the odd part of the electric field along the $x$-axis.
Decomposing the parity operator into the HG basis we find all \tem{n,m} modes that have an odd mode order index $n$ for the x-axis pick up an additional minus sign.

For the LCT we can simplify this analysis since the parity operation $\mathcal{P}$ is a subset of the LCT that can be represented by an ABCD matrix, namely the negative identity matrix
\begin{align}
    \mathbf{M}_\mathcal{P} = \begin{bmatrix*}
        -1 & \phantom{-}0\,\, \\
        \phantom{-}0 & -1\,\,
    \end{bmatrix*},
\end{align}
which can be interpreted as the reflection off of a flat mirror.
It should be noted that in separable 2D models the parity operation should only be included in the $x$-axis ABCD matrix.
So for example the ABCD matrices for reflecting off of a curved mirror in the $x$ and $y$ axis respectively are given by
\begin{align}
    \mathbf{M}_x = \begin{bmatrix*}
        -1 & \phantom{-}0\,\, \\
        \frac{2}{R_x} & -1\,\,
    \end{bmatrix*}, && \mathbf{M}_y = \begin{bmatrix*}
        1 & 0 \\
        \frac{-2}{R_y} & 1
    \end{bmatrix*},
\end{align}
where $R_x$ and $R_y$ are the radii of curvature of the mirror in the $x$ and $y$ axes respectively.

This approach of encoding the parity transformations with the ABCD matrix allows \eqref{eq:as2} to be valid for both linear and ring cavities. 
It has also previously been justified by Siegman based on purely geometrical arguments in figure 15.8 of~\cite{siegman_15_1986}, and by Arai~\cite{arai_accumulated_2013} for computing the transverse mode splitting in ring cavities.

It should be noted that encoding the parity operator into the ABCD matrix introduces 90$^\circ$ of phase on reflection to the entire electric field from the $\sqrt{D}$ term in \eqref{eq:continuous_LCT_def}.
This affects the phase relationship between reflection and transmission that is needed to maintain conservation of energy (section 2.4 in~\cite{bond_interferometer_2016}).
However, it can be trivially accounted for by defining all scalar amplitude reflectivies to have an additional -90$^\circ$ of phase to counteract the 90$^\circ$ from the LCT parity operation.

\subsection{Results}

\begin{figure}[ht]
	\centering
    \includegraphics[width=\linewidth]{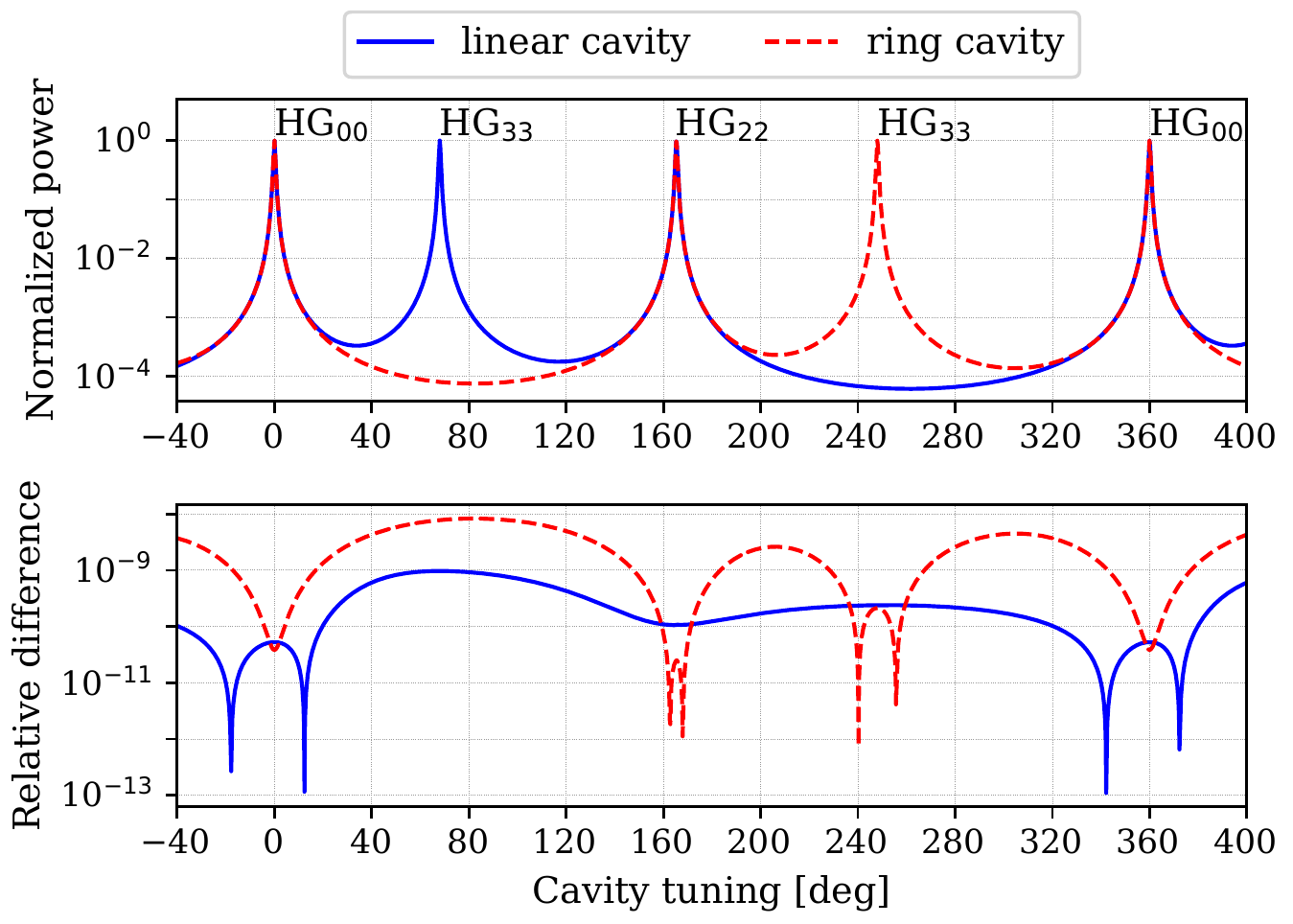}
     \caption{The upper plot shows the calculated circulating power inside a linear (solid blue) and triangular (red dashed) cavity with infinite mirrors versus cavity tuning $\phi$.
    The free spectral range (FSR) and round trip Gouy phase identical to a 4~km aLIGO arm cavity.
    The power is normalized so that on-resonance the circulating power is 1.
    The LCT model uses a $200 \times 200$ grid.
    The circulating power computed by both models is indistinguishable by eye and so only the LCT solution is shown in the upper plot.
    The relative difference between the LCT and \textsc{Finesse} models are shown in the lower plot.
     }
    \label{fig:lct_cavity_scan}
\end{figure}

The normalized circulating power predicted by \eqref{eq:as2} as the round trip phase, $\phi$, is varied is plotted in figure~\ref{fig:lct_cavity_scan} for a linear and ring cavity.
The incident field consists of the sum of equal power \tem{00}, \tem{22}, and \tem{33} modes.
Note that for the \tem{33} the resonance phase differs in the linear and ring cavities by 180$^\circ$ due to the overall parity operator in the round trip in a ring cavity with an odd number of mirrors.

The ciruclating power as computed in the LCT model is compared to an equivalent HG modal model calculated using the \textsc{Finesse} optical modelling package~\cite{freise_frequency-domain_2004,freise_finesse_2013,brown_finesse_2014,brown_pykat_2020}.
Figure~\ref{fig:lct_cavity_scan} shows that the LCT and HG model agree to at least $10^{-8}$ of the circulating power in the test case shown.

The difference between the two models seen in figure~\ref{fig:lct_cavity_scan} can be attributed entirely due to the limited resolution in the LCT grid as the HG model in this case corresponds to the complete analytic solution.

\section{Linear cavity with apertures and mirror defects}\label{sec:apertured_cavity}
We have only considered optical models that are able to be completely described by ABCD matrices and hence use mirrors of infinite extent.

To incorporate finite aperture mirrors into the DLCT framework we need to formulate the action of an aperture as a linear operator.
For an electric field $\mathbf{X}$ sampled on a 2D $N \times N$ grid, apertures (and most mirror surface maps) become diagonal $N^2 \times N^2$ matrices when implemented as a linear operator.
The off-diagonal elements become non-zero when propagating the beam away from the mirror and so in general all $N^2 \times N^2$ elements are non-zero in any optical model.
This poses a technical challenge as the computer storage requirements of an $N^2 \times N^2$ matrix are prohibitively large for current consumer hardware for $N>200$.
Specifically for 128-bit complex floating point numbers, the amount of memory required to store an $N^2 \times N^2$ matrix is $N^4 \times 1.6 \times 10^{-8}$~GB.
This $\mathcal{O}(N^4)$ scaling in storage requirements has been documented as a primary bottleneck in similar linear operator based approaches~\cite{new_computing_2017}.

To reduce this $\mathcal{O}(N^4)$ space complexity we loosen the requirement that a linear operator needs to be represented by an explicit matrix by recognizing that a linear operator in the space spanned by DLCTs and mirror maps can be represented by an $N^2 \times N^2$ \textit{structured matrix} $\mathbf{A}$, where the $\textit{structure}$ allows the matrix vector multiplication $\mathbf{Ax}$ to be computed using only $N \times N$ operations, much faster than what is possible with a general dense $N^2 \times N^2$ matrix.
This reduces the space complexity from $\mathcal{O}(N^4)$ to $\mathcal{O}(N^2)$ at the cost of not having the explicit matrix elements of $\mathbf{A}$, which restricts the kind of algorithms that can be used to compute the circulating field to a subset of iterative linear algebra algorithms.

To compute the circulating field with apertures we first consider the general operator solution in \eqref{eq:cavity_graph_circulating_field}.
Unlike before where each operator was given by a DLCT matrix, we now define the matrix-vector multiplication procedure of each operator acting on some 2D input array $\mathbf{X}$ in the following way
\begin{align}
    \mathcal{T}_{3 \gets 1} [\mathbf{X}] &= \iu t_1 \mathbf{T}_1 \circ \mathbf{X}\\
    \mathcal{D}_{5\gets3}[\mathbf{X}] = \mathcal{D}_{4\gets6}[\mathbf{X}] &= e^{i \phi/2} \dlct{\mathbf{M}_d} \mathbf{X} (\dlct{\mathbf{M}_d})^T\\
    \mathcal{R}_{6\gets5} [\mathbf{X}] &= r_2 \mathbf{R}_2 \circ \mathbf{X}\\
    \mathcal{R}_{3\gets4} [\mathbf{X}] &= r_1 \mathbf{R}_1 \circ \mathbf{X} \label{eq:ETM_2D_reflection_map}
\end{align}
where $\circ$ is the Hadamard product (a.k.a. element-by-element array product), $\mathbf{T}_1$ is the transmission map of the input mirror $\mathbf{R}_1$ and $\mathbf{R}_2$ are reflection maps of the input and output mirror respectively.
There are no structural requirements on these reflection and transmission maps and as such they can contain arbitrary apertures and mirror deformations.
$\dlct{\mathbf{M}_d}$ is the 1D DLCT for a Fresnel diffraction of $d$ meters.
This completes our specification of all of the operators in a 2D linear cavity on an $N \times N$ grid with arbitrary apertures and mirror distortions using only $N \times N$ array and matrix multiplication operations.

To solve \eqref{eq:cavity_graph_circulating_field} it is more convenient to rewrite it into a familiar form
\begin{align}
    \mathcal{A} [\mathbf{X}] = \mathbf{B} \label{eq:structured_mat_vec_eq}
\end{align}
where we relabel the circulating field $\mathbf{E}_3$ to $\mathbf{X}$, $\mathbf{B} = \mathcal{T}_{3 \gets 1} [\mathbf{E}_1]$, and $\mathcal{A}$ is a procedure given by
\begin{align}
    \mathcal{A}[\mathbf{X}] = \mathcal{I}[\mathbf{X}] - \mathcal{R}_{3\gets4}\mathcal{D}_{4\gets6}\mathcal{R}_{6\gets5}\mathcal{D}_{5\gets3}[\mathbf{X}].
    \label{eq:structured_circ_operator}
\end{align}
\eqref{eq:structured_mat_vec_eq} can be interpreted as a linear equation in $\mathbf{X}$ as $\mathcal{A}$ is made up of only linear operations it itself must be a linear operator where $\mathcal{A}[\mathbf{x} + \mathbf{y}] = \mathcal{A}[\mathbf{x}] + \mathcal{A}[\mathbf{y}]$ for all $\mathbf{x}$ and $\mathbf{y}$.

Despite lacking the explicit matrix elements of the linear operator $\mathcal{A}$, the procedure $\mathcal{A}[\mathbf{X}]$ gives us the ability to compute its matrix-vector product in an efficient way.
A number of algorithms are available to solve \eqref{eq:structured_mat_vec_eq} using only the procedure that computes the matrix-vector product $\mathcal{A}[\mathbf{X}]$. 
These algorithms typically employ the use of a Krylov subspace such as GMRES~\cite{saad_gmres_1986}, and BICGSTAB~\cite{van_der_vorst_bi-cgstab_1992}, both of which have easy to use python wrappers available in scipy~\cite{scipy_10_contributors_scipy_2020}.

\subsection{Example: aLIGO linear cavity with finite-aperture mirrors}
\label{sec:apertured_aligo}

\begin{figure}[!ht]
	\centering
	\includegraphics[width=\linewidth]{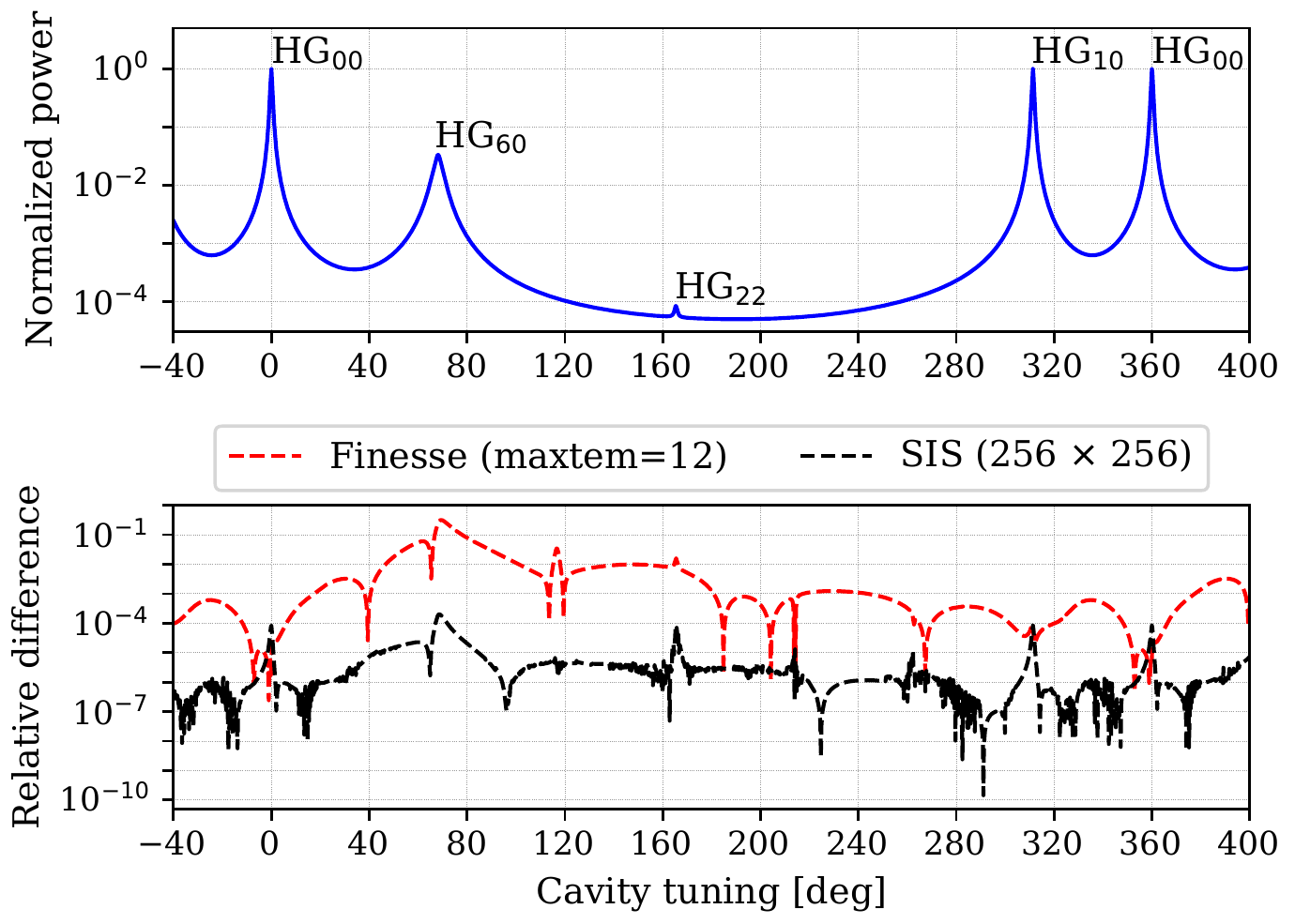}
	\caption{The upper plot is the circulating power inside an aLIGO arm cavity with 34~cm diameter mirrors as a function of cavity tuning $\phi$.
    The difference between each model is indistingiushable by eye and so only the LCT circulating power is plotted.
    The lower plot shows the relative difference between the LCT model and an HG model Finesse (dashed red) and the SIS~\cite{yamamoto_sis_2008} FFT model (dashed black).
	The \textit{maxtem N} parameter tells \textsc{Finesse} to only model the \tem{nm} modes that have their mode order $n + m \le N$. 
	Both the LCT and FFT model use the same 256 $\times$ 256 grid.
	}
	\label{fig:apertured_cavity_scan}
\end{figure}

For the numerical comparison we once again perform a cavity scan on the aLIGO arm cavity as in figure~\ref{fig:lct_cavity_scan}, except now we add 34~cm circular apertures to both mirrors in the arm cavity, corresponding to the diameter of the aLIGO arm cavity mirrors.
We change our incident beam to be composed of equal parts \tem{00}, \tem{10}, and \tem{60}.
The resulting cavity scan is shown in figure~\ref{fig:apertured_cavity_scan}.

The \tem{10} and \tem{60} modes that were chosen primarily for their difference in physical extent. 
The larger \tem{60} interacts more strongly with the aperture than the \tem{10} and is therefore expected to have a higher clipping loss.
This can be seen in figure~\ref{fig:apertured_cavity_scan} where the \tem{60} resonance is both smaller and wider than the \tem{10} resonance.
Additionally, a small resonance can be seen 165 degrees that corresponds to a fourth order mode.
This resonance is explained by the mode scattering of the \tem{60} by the aperture to produce fourth order modes.

The LCT prediction is compared in figure~\ref{fig:apertured_cavity_scan} (lower) with the modal-based: Finesse, and an FFT-based model: SIS, which is typically used for modelling high spatial frequency effects in optical cavities.
While the LCT and SIS models agree to within about 1~ppm, the disagreement between Finesse and LCT is several orders of magnitude worse, particularly at the heavily-clipped \tem{60} mode.

This disagreement is perhaps expected as in \textsc{Finesse} an aperture is represented by a scattering matrix where each element describes how the amplitude of each apertured HG mode couples to each HG modes.
The elements of this scattering matrix are given by overlap integrals, which are computed numerically.
The accuracy of an HG model with apertures is then dependent on the numerical accuracy of the scattering matrix element integration, and the number of HG modes used.
However, it is practically infeasible to use enough modes to accurately model hard aperture effects~\cite{bond_interferometer_2016}.

\subsection{Case study: aLIGO point absorbers}

Example \ref{sec:apertured_aligo} assumed that the reflection maps for the mirrors described a uniform reflectivity, spherical mirror with a finite circular clear aperture.
Here, we consider the effect of a point-like absorbing defect on the mirror surface, such as those discovered in the aLIGO arm-cavity mirrors~\cite{the_ligo_scientific_collaboration_advanced_2015}.
These point absorbers caused a large decrese in the power recycling gain (PRG) as the power stored in the arms increased~\cite{buikema_sensitivity_2020}.
An analysis of this effect has recently been published by Brooks et~al.~\cite{brooks_point_2021}

The reflection map for a mirror $m$; $\mathbf{R}_m$ is given by
\begin{align}
    (\mathbf{R}_m)_{ij} = \begin{dcases}
        0 \hspace{26ex} \text{if $\sqrt{x_j^2 + y_i^2} \ge \frac{D_m}{2}$}\\
        r_m \times \exp\left[-\frac{\iu 4 \pi}{\lambda} \big[(\textbf{S}_m)_{ij} + (\mathbf{C}_m)_{ij} \big] \right] \hspace{5ex} \text{otherwise,}\\
    \end{dcases}
\end{align}
where $r_m$ is the amplitude reflection coefficient, $D_m$ is the diameter of the mirror,
\begin{align}
    (\mathbf{C}_m)_{ij} = -\frac{x_j^2}{2 R_{m,x}} -\frac{y_i^2}{2 R_{m,y}}
\end{align}
is the curvature height map and $R_{m,x}$, and $R_{m,y}$ are the radii of curvature along the x, and y-axis repsectively.
The point absorber height map $\mathbf{S}_m$ is derived using Eq.~B8 and B9 in~\cite{brooks_point_2021}.
\begin{align}
    \left(\mathbf{S}_m\right)_{ij} = \begin{dcases}
        -\frac{\alpha P_\text{abs}}{2 \pi \kappa}\left(\frac{r_{ij}}{4\omega}\right)^2 & r_{ij} \leq \omega\\
        \frac{\alpha P_\text{abs}}{2 \pi \kappa}\left(-\frac{1}{2} + \log\left[ \frac{\omega \left( h_m^2 + r_{ij}^2 \right)}{r_{ij} \left( h_m^2 + \omega^2 \right)} \right]\right) & r_{ij} > \omega
    \end{dcases}
\end{align}
where 
\begin{align}
    r_{ij} = \sqrt{\Big(x_j - x_0\Big)^2 + \Big(y_i - y_0\Big)^2}
\end{align}
and $x_0$, $y_0$ is the location of the point absorber, $\omega$ is the radius of the point absorber, $h_m$ is the thickness of the mirror, $P_\text{abs}$ is the absorbed power, and $\frac{\alpha}{2 \pi \kappa} \approx 6.3\times10^{-8}$~m/W is a scaling factor derived from the thermal expansion coefficient $\alpha$ and thermal conductivity $\kappa$ of fused silica.

\begin{figure}[!ht]
	\centering
    \includegraphics[width=\linewidth]{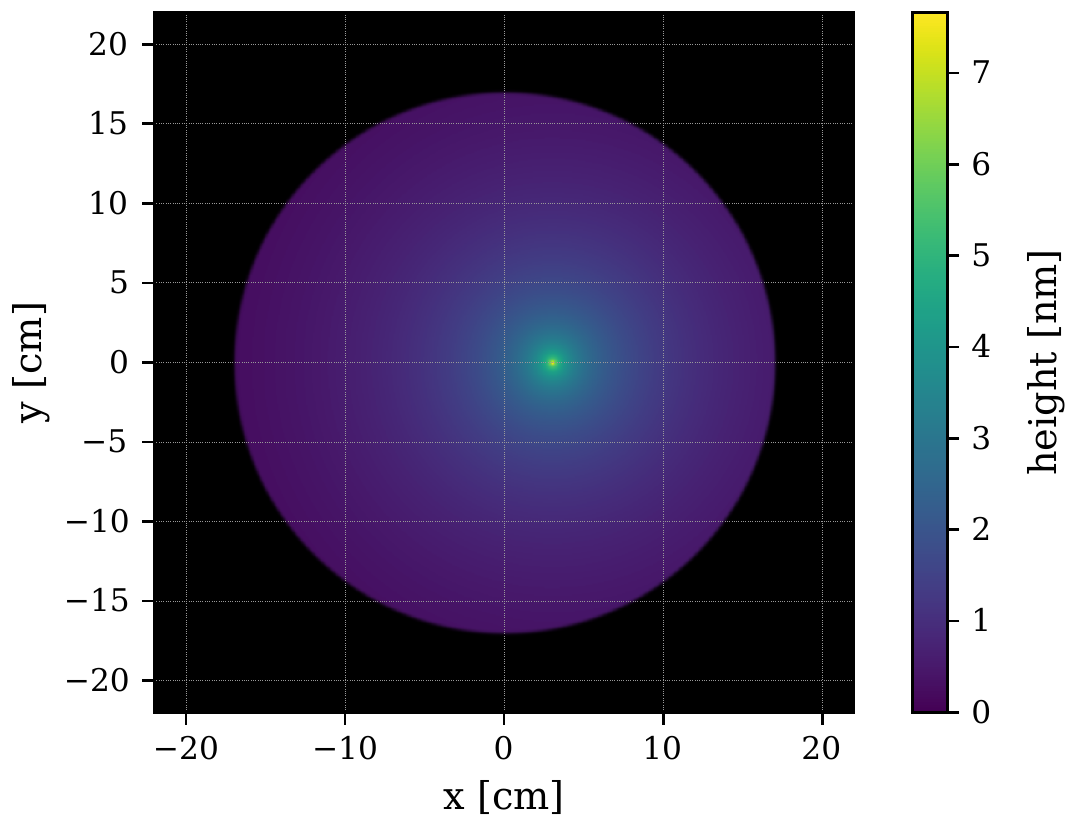}
     \caption{A height map of a 34~cm mirror with a 40~$\mu$m diameter point absorber with 20 mW of absorbed power offset 3 cm in the x-axis from the center of the optic.}
    \label{fig:point_absorber_map}
\end{figure}

\begin{figure}[!ht]
	\centering
    \includegraphics[width=\linewidth]{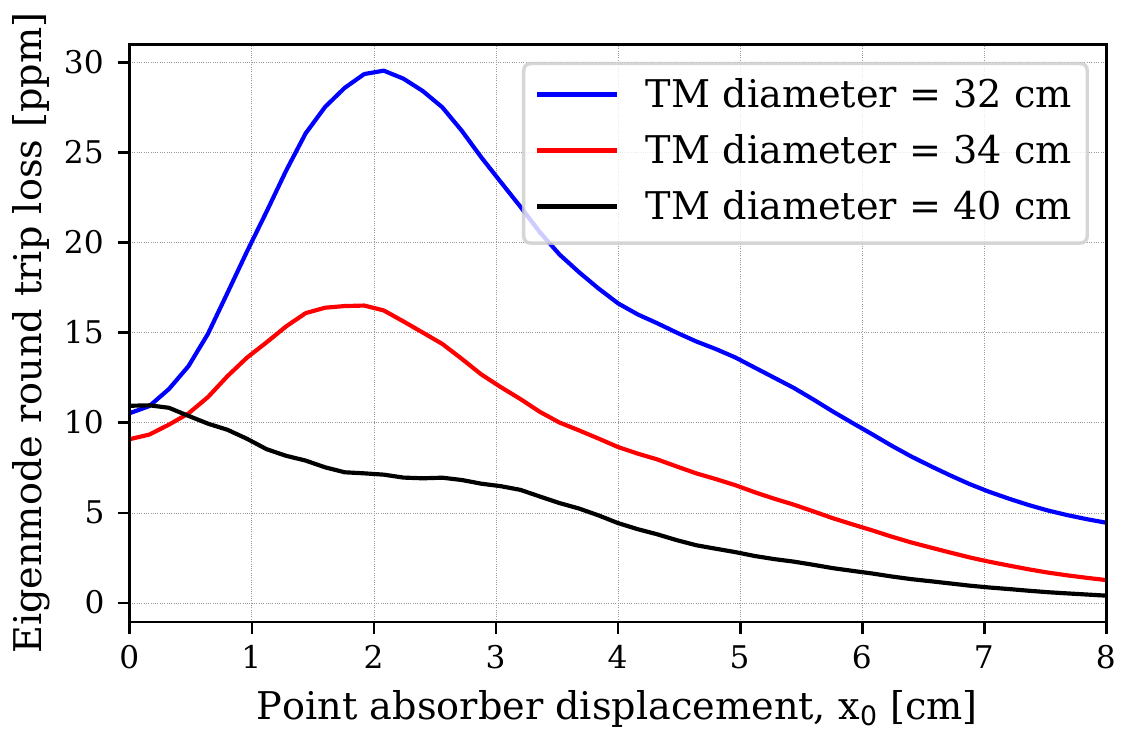}
     \caption{A plot of round trip loss for the lowest order mode in the aLIGO arm cavity versus point absorber position along the x-axis for three different arm cavity test mass (TM) mirror diameters: 32~cm (blue), 34~cm (red), and 40~cm (black).
     The loss is explicitly computed as $1-|\gamma_0|^2$.}
    \label{fig:point_absorber_postion_scan}
\end{figure}

A map of a surface bump due to a 40~$\mu$m diameter point absorber offset by 3~cm from the center of a 34~cm diameter, 17~cm thick mirror and 20~mW of asbsorbed power is shown in figure~\ref{fig:point_absorber_map}.
The eigenmodes and eigenvalues of  \eqref{eq:structured_circ_operator} were computed using the Arnoldi method in ARPACK~\cite{lehoucq_arpack_1998} fpr a range of absorbed powers and defect locations. While the shape of the lowest-order eigenmode is almost identical to the \tem{00} mode, its eigenvalue varies significantly with point absorber position and mirror diameter as shown in figure~\ref{fig:point_absorber_postion_scan}.

We can compute the decrease in PRG that would result from this point absorber loss by using Eq.~6 in \cite{izumi_advanced_2016}
\begin{align}
	\text{PRG} &= \left(\frac{t_p}{1 - r_p r_a}\right)^2
\end{align}
where $t_p$ and $r_p$ are the amplitude transmission and reflection coefficients of the power recycling mirror (PRM) respectively. In this equation and following two, $r = \sqrt{1-t^2}$.
We define the effective arm cavity amplitude reflection coefficient $r_a$ as
\begin{align}
    r_a &= \frac{1}{r_i}\left( \frac{g_\text{rt}-r_i^2}{1-g_\text{rt}} \right)
\end{align}
where  $t_i$ and $r_i$ are the amplitude transmission and reflection coefficients of the input test mass (ITM) and $g_\text{rt}$ is the round trip arm gain on resonance, given by
\begin{align}
    g_\text{rt} &= r_i r_e |\gamma_{0}| \sqrt{1-\Gamma}
\end{align}
where $r_e$  is the amplitude reflectivity of the end test mass (ETM) mirror, $\gamma_{0}$ is the round trip eigenvalue of the $u_0$ arm cavity eigenmode, and $\Gamma$ is an additional round trip loss factor.
Taking the absolute value of the eigenvalue $\gamma_{0}$ is effectively equivalent to making the arm cavity resonant for the $u_0$ eigenmode.

A summary of the necessary constant parameters needed to compute the PRG are given in table~\ref{tab:PRG_parameters}.
\begin{table}[H]
	\centering
	\begin{tabular}{|c|c|c|c|}
		\hline
		 $t_p^2$ & $t_i^2$ & $t_e^2$ & $\Gamma$ \\
		 \hline
         0.03 & 0.014 & $5\times10^{-6}$ & $60\times10^{-6}$\\
		 \hline
	\end{tabular}
\caption{Relevant aLIGO parameters for computing the PRG.}\label{tab:PRG_parameters}
\end{table}

\begin{figure}[!ht]
	\centering
    \includegraphics[width=\linewidth]{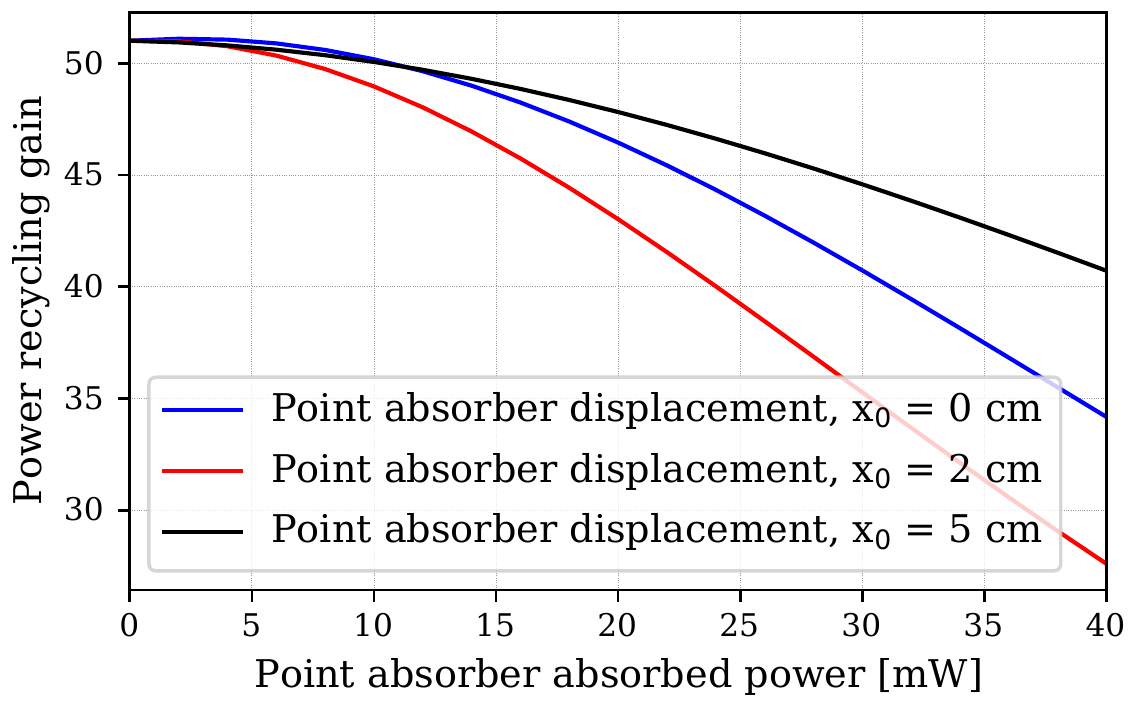}
     \caption{A plot of the PRG decreasing with increased power absorbed by a 40~$\mu$m point absorber located at 0~cm (blue), 2~cm (red), and 5~cm (black) away from the center of the optic.}
    \label{fig:point_absorber_power_scan}
\end{figure}

In this study we assume the point absorber has a negligible effect on all previously mentioned parameters with the exception of $\gamma_{0}$.
Computing the PRG as a function of point absorber absorbed power results for three different absorber positions is shown in figure~\ref{fig:point_absorber_power_scan}.
The additional round trip loss $\Gamma$ was set to 60~ppm to match the experimentally observed PRG at low power. We assume that for less than 40~mW of absorbed power the height map is linearly proportional to the absorbed power.
This is to be compared to figure~4 in~\cite{brooks_point_2021}, which shows a similar trend for PRG decreasing as a function of circulating power in the arm cavity. 

Perhaps surprisingly the loss increases initially as the point absorber moves away from the center of the optic, peaking at about 2~cm where afterwards it monotonically decreases. 
This is better shown in figure~\ref{fig:point_absorber_postion_scan} where the point absorber power is kept constant at 20~mW as it is moved across the optic.
This loss peak at 2~cm is very sensitive to the arm cavity mirror size. 

\begin{figure}[!ht]
	\centering
    \includegraphics[width=\linewidth]{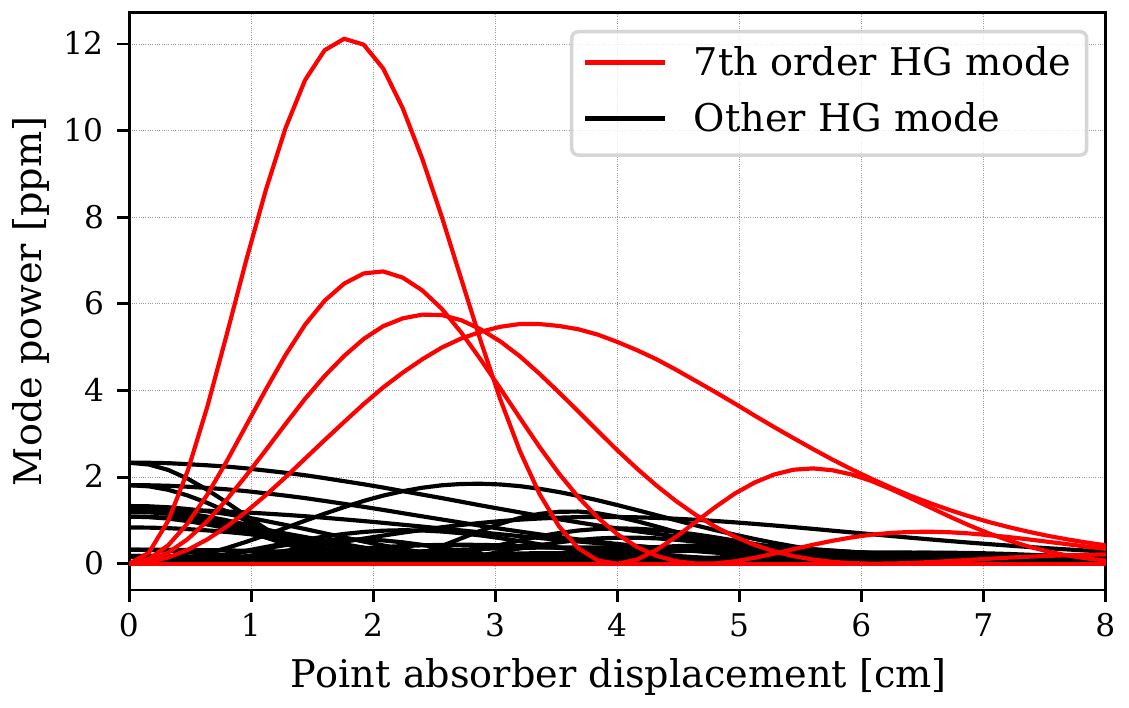}
     \caption{An HG decomposition of the arm cavity circulating field with 34 cm mirrors.
     The seventh order modes are highlighted in red. The power in all \tem{n,m} with $n+m \leq 10$ are plotted with the exception of the \tem{00} mode.}
    \label{fig:point_absorber_HG_decomp}
\end{figure}

The explanation for the peak in arm cavity loss at 2~cm seen in figure~\ref{fig:point_absorber_postion_scan} is that the point absorber scattering maximizes scattering the \tem{00} into 7th order modes.
The 7th order modes resonate nearly at the same phase as the \tem{00} and so are partially resonantly enhanced when the arm cavity is locked onto the \tem{00}.
This can be most clearly seen by performing an HG decomposition on the arm cavity circulating field as was done in figure~\ref{fig:point_absorber_HG_decomp}.

The explanation behind the dependence on mirror size then follows as the larger 7th order modes interact strongly with the edges of the hard aperture.
Decreasing the size of the mirrors increases the loss on the 7th order modes, thus broadening their resonance, allowing for more resonant enhancement from the \tem{00}.
Conversely, increasing the size of the mirrors narrows the 7th order resonance, decreasing the enhancement from the \tem{00}.

A similar conclusion about the contribution of 7th order modes to the loss was made by Brooks et al.~\cite{brooks_point_2021}, though following a different formalism. This case study demonstrates that a combination of apertures, reflection maps, and resonant conditions must be modelled sufficiently to see these HOM loss effects.

\section{Conclusion}
We have presented an introduction to LCTs and their application to free-space wave propagation through first order optical systems.
A new application of the LCT to efficiently solving for the circulating field in a Fabry-Perot cavity using general purpose sparse linear algebra algorithms was demonstrated.
The LCT method shows good numerical agreement with analytical solutions and established FFT-models, as well as reproducing results from recent aLIGO point absorber models. This highlights how such modeling tools can be used in designing future detectors to avoid such problems.
Future work will need to extend the LCT method to coupled cavities to allow us to model complete interferometric systems. Further work is also needed to identify which linear algebra methods optimally solve the LCT matrices as well as packaging it up into a usable tool, such as \textsc{Finesse} or SIS.

\appendix
\section{Metaplectic sign ambiguity}\label{sec:metaplectic_sign}
The LCT has some implications when calculating the accumulated Gouy phase through an optical system---which is due to the LCT being a faithful representation of the metaplectic group. Here we find a new formula for the 1D accumulated Gouy phase, which can also be used to compute the accumulated Gouy phase in a simple astigmatic 2D optical system.

The ABCD matrices discussed here form a group known as the symplectic group $\text{Sp}_2(\mathbb{R})$~\cite{wolf_geometric_2004}, which are all $2\times2$ matrices with unit determinant.
There are 2 distinct LCTs for every ABCD matrix because the metaplectic group $\text{Mp}_2(\mathbb{R})$ is a double cover of the symplectic group $\text{Sp}_2(\mathbb{R})$.
The additional parameter that allows us to uniquely specify an LCT is called the metaplectic sign~\cite{wolf_geometric_2004}.

The metaplectic sign manifests in optics as a sign ambiguity in the Gouy phase accumulated by a beam propagating through an ABCD optical system.
This sign ambiguity disappears in 2D optical systems exhibiting cylindrical symmetry due to both vertical and horizontal having the same metaplectic sign, which end up cancelling yielding an overall expression for accumulated Gouy phase that has no sign ambiguity~\cite{erden_accumulated_1997,siegman_20_1986}.

The accumulated Gouy phase for a Gaussian beam with beam parameter $q$ through an ABCD system in two dimensions with cylindrical symmetry is
\begin{equation}
    \text{exp}(\iu \Psi) = \text{exp}(\iu 2 \psi) = \frac{A + B/q^*}{|A + B/q^*|}
    \label{eq:accum_gouy_2d}
\end{equation}
and in one dimension 
\begin{equation}
    \text{exp}(\iu \psi) = \pm \sqrt{\frac{A + B/q^*}{|A + B/q^*|}}
    \label{eq:accum_gouy_1d}
\end{equation}
where the $\pm$ sign, which we identify as the \textit{metaplectic sign} is a real sign ambiguity that is present in the 1D accumulated Gouy phase for an arbitrary ABCD matrix.

An algorithm to determine the metaplectic sign for LCTs to the best of our knowledge was first presented by Littlejohn (appendix A of~\cite{littlejohn_semiclassical_1986}, and \cite{littlejohn_new_1987}) and has been subsequently used by Lopez~\cite{lopez_pseudo-differential_2019}.
Their algorithm computes and tracks a winding number from the multiplication of ABCD matrices, which is then used to determine the metaplectic sign.
We propose a modification to this algorithm that bypasses the need to track the winding number and instead tracks the metaplectic sign directly.
We use this algorithm to define a group action, which is an extension of regular matrix multiplication between ABCD matrices paired with a metaplectic sign, which we call a \textit{metaplectic matrix}.

\subsection{Metaplectic Matrices}

A metaplectic matrix $\mathfrak{M} \in \text{Mp}_2(\mathbb{R})$ is a composite object that can be represented as a tuple/ordered set containing a symplectic matrix $\mathbf{M} \in \text{Sp}_2(\mathbb{R})$ and a binary sign $\sigma \in \{-1, 1\}$, which is the metaplectic sign.
\begin{equation}
    \mathfrak{M} = \{\mathbf{M}, \sigma\}.
\end{equation}
The multiplication law between two metaplectic matrices is then defined as
\begin{equation}
    \mathfrak{M}_3 = \mathfrak{M}_2 \mathfrak{M}_1 = \{\mathbf{M}_2 \mathbf{M}_1, \sigma_3\},
\end{equation}
where the new sign $\sigma_3$ is given by
\begin{equation}
    \sigma_3 = ( \sigma_2 \times \sigma_1 ) \times (-1)^{\rho(\mathbf{M}_2, \mathbf{M}_1)}
\end{equation}
where \hbox{$\rho(\mathbf{M}_2, \mathbf{M}_1) \in \{0, 1\}$} is given by
\begin{equation}
    \rho(\mathbf{M}_2, \mathbf{M}_1) = [\theta(\mathbf{M}_2) < 0] \oplus [\theta(\mathbf{M}_2 \mathbf{M}_1) < \theta(\mathbf{M}_1)] \label{eq:metaplectic_sign_flip}
\end{equation}
where $\oplus$ is a logical exclusive or (XOR), and $-\pi < \theta(\mathbf{M}) \leq \pi$ is a metaplectic phase associated with an ABCD matrix given by
\begin{equation}
    \theta(\mathbf{M}) = \text{arg}(A + \iu B).
\end{equation}
We adopt the convention of the boolean True and False being represented by 1 and 0 respectively.

\eqref{eq:metaplectic_sign_flip} was derived from logical expressions of the following kind: if the metaplectic phase is decreasing $[\theta(\mathbf{M}_2) < 0]$, and if the output metaplectic phase is less than the input metaplectic phase $[\theta(\mathbf{M}_2 \mathbf{M}_1) < \theta(\mathbf{M}_1)]$, then the metaplectic sign hasn't flipped $[\rho(\mathbf{M}_2, \mathbf{M}_1) = 0]$.
Enumerating all possible combinations of True and False for the conditions in \eqref{eq:metaplectic_sign_flip} reveals that the truth table for determining if a metaplectic sign flip has occurred is equivalent to an XOR of the two conditions.

The 1D Gouy phase of a Gaussian beam with beam parameter $q$ through a metaplectic ABCD matrix $\mathfrak{M} = (\mathbf{M}, \sigma)$ is then
\begin{equation}
    \text{exp}(\iu \psi) = \sigma \sqrt{\frac{A + B/q^*}{|A + B/q^*|}}.
    \label{eq:metaplectic_gouy_1d}
\end{equation}
With this we can define the Gouy phase of a Gaussian beam through a 2D optical system with simple astigmatism described with two metaplectic matrices $\mathfrak{M}_x$ and $\mathfrak{M}_y$ for the $x$ and $y$ axes respectively as 
\begin{equation}
    \text{exp}(\iu \Psi) = \sigma_x \sigma_y \exp \!\big[\iu (\psi_x + \psi_y )\big]
    \label{eq:metaplectic_gouy_2d}
\end{equation}
which reduces to \eqref{eq:accum_gouy_2d} to in the case of cylindrical symmetry, where $\mathfrak{M}_x = \mathfrak{M}_y$.

It should be noted that the expressions for Gouy phase in \eqref{eq:metaplectic_gouy_1d} and \eqref{eq:metaplectic_gouy_2d} are not analytic (i.e. they don't have Taylor series) due to the discrete nature of the metaplectic sign $\sigma \in \{-1, 1\}$.
We believe that it is likely impossible to derive an analytical expression for 1D Gouy phase that uses a single ABCD matrix to represent an arbitrary optical system.
A similar statement has been made by other authors for 2D Gouy phase in general astigmatic optical systems~\cite{habraken_geometric_2010}.




\bibliography{main}

\end{document}